\documentclass[runningheads]{llncs}
\usepackage{mathtools} 

\usepackage{array}
\usepackage{graphicx}
\usepackage{amsmath,amssymb}
\usepackage[misc]{ifsym}
\usepackage{capt-of}
\usepackage{hyperref}
\usepackage{hhline,boldline}
\usepackage{adjustbox}
\usepackage[caption=false]{subfig}
\hypersetup{
    colorlinks=true,
    linkcolor=blue,
    filecolor=magenta,      
    urlcolor=blue,
    citecolor = blue
}
\usepackage{pdfpages}

\begin{document}

\title{Generation of 3D Brain MRI Using Auto-Encoding Generative Adversarial Networks}

\titlerunning{Generation of 3D Brain MRI Using Auto-Encoding GAN}

\author{Gihyun Kwon\inst{1}\and Chihye Han\inst{1}\and
Dae-shik Kim\inst{1}$^{\textrm{(\Letter)}}$}
\authorrunning{G. Kwon et al.}
\institute{Korea Advanced Institute of Science and Technology, Daejeon, South Korea 
\email{\{cyclomon,ckelseyhan,daeshik\}@kaist.ac.kr}}

\maketitle
\begin{abstract}
As deep learning is showing unprecedented success in medical image analysis tasks, the lack of sufficient medical data is emerging as a critical problem. While recent attempts to solve the limited data problem using Generative Adversarial Networks (GAN) have been successful in generating realistic images with diversity, most of them are based on image-to-image translation and thus require extensive datasets from different domains. Here, we propose a novel model that can successfully generate 3D brain MRI data from random vectors by learning the data distribution. Our 3D GAN model solves both image blurriness and mode collapse problems by leveraging $\alpha$-GAN that combines the advantages of Variational Auto-Encoder (VAE) and GAN with an additional code discriminator network. We also use the Wasserstein GAN with Gradient Penalty (WGAN-GP) loss to lower the training instability. To demonstrate the effectiveness of our model, we generate new images of normal brain MRI and show that our model outperforms baseline models in both quantitative and qualitative measurements. We also train the model to synthesize brain disorder MRI data to demonstrate the wide applicability of our model. Our results suggest that the proposed model can successfully generate various types and modalities of 3D whole brain volumes from a small set of training data.

\end{abstract}

\section{Introduction}
Recent deep neural networks, especially Convolutional Neural Networks (CNN), have shown outstanding performance in various computer vision tasks such as classification and segmentation based on the availability of large data. Along with these achievements, areas of medical image analysis including disease diagnosis and lesion detection have also made a remarkable breakthrough using CNNs. However, training CNN-based models requires a large set of medical image data~\cite{Ravi_2017}, which are laborious and costly to obtain. Conventional geometric transformation methods (e.g. flip, rotation) can be used to augment training data, but their outputs are highly dependent on the original data.

One actively explored solution for the data deficit problem is Generative Adversarial Networks (GAN)~\cite{GAN}, which have succeeded in the computer vision domain by generating natural images that are realistic but different from the original. Two major approaches exist to medical image generation using GANs: image-to-image translation and generating images from random distribution. The former has been extensively explored in various contexts such as T1-T2 cross-modal synthesis of 2D slices~\cite{Dar_2019}, modality conversion from T1 to FLAIR in 3D volumes~\cite{Yu_2018}, and 3D brain tumor MRI synthesis from normal control ~\cite{Shin_2018}. In this approach, training is relatively easy as it is done with the guidance of another dataset, and the quality of generated images is comparable to that of real images. However, it requires extensive training data, and the generated output rely on the attributes of the original data such as shape. In the latter, Han et al.~\cite{Han_2018} and Bermudez et al.~\cite{Bermudez_2018} generated realistic 2D tumor and normal MRI slices from random variables, respectively. This method can generate completely new images with more variability by learning the data distribution itself. However, it is rarely attempted due to difficulties in stabilizing training, which limited previous works to generating 2-dimensional slices.

In this paper, we propose a 3D GAN model that successfully generates 3D brain MRI from random vectors. We adapt the structure of $\alpha$-GAN~\cite{alphaGAN} for 3D image generation, which addresses mode collapse and image blurriness by introducing additional auto-encoder and code discriminator networks on top of the existing generator and discriminator. We also utilize Wasserstein GAN with Gradient Penalty (WGAN-GP)~\cite{WGAN-GP} loss functions to prevent unstable training. To the best of our knowledge, this work is the first attempt to generate completely new 3D brain MRI from random distribution. To demonstrate the versatility of our model, we train the model with brain tumor and stroke lesion MRI and show that it can generate realistic 3D whole brain images of multiple types (e.g. normal or diseased) and modalities (e.g. T2 or FLAIR). The proposed model is expected to be widely applicable for medical image analysis tasks such as disease diagnosis by enabling use of cross-modal information, and especially useful for rare diseases, as it requires a small amount of data.

\section{Methods}
\noindent
\textbf{Model Architecture.} 
The main challenge in 3D generation is that the mode collapse problem, where GANs produce only a limited variety of images, becomes more severe as the complexity of the task increases abruptly going from 2D to 3D generation. A natural alternative is to use Variational Auto-Encoder (VAE)~\cite{VAE}, which is free from mode collapse but outputs are characterized with blurriness. In order to effectively address the problems of both mode collapse of GANs and blurriness of VAEs, we use $\alpha$-GAN~\cite{alphaGAN}, a solution born by combining both models. $\alpha$-GAN proposes a code discriminator network $C$ that replaces variational inference in VAE, a process in which the posterior from the encoder $z_e \sim q_e(z|x)$ is explicitly set as a specific distribution such as Gaussian and matched to the random prior $z_r \sim P(z)$ with the reparametrization trick. Since this process has a different objective from GANs, $\alpha$-GAN transforms this inference in a more GAN-like fashion: By treating the encoder output $z_e \sim q_e(z|x)$ as fake and $z_r$ as real, the encoder and the code discriminator play an adversarial game. When $C$ is not able to discriminate both, the posterior and the prior are perfectly matched. Thus, the posterior probability can be estimated implicitly. Under the unified objective of VAE and GAN, benefits of both models can be exploited, and the generation performance improves. 
\begin{figure}[t]
    \centering
    \includegraphics[width=\textwidth]{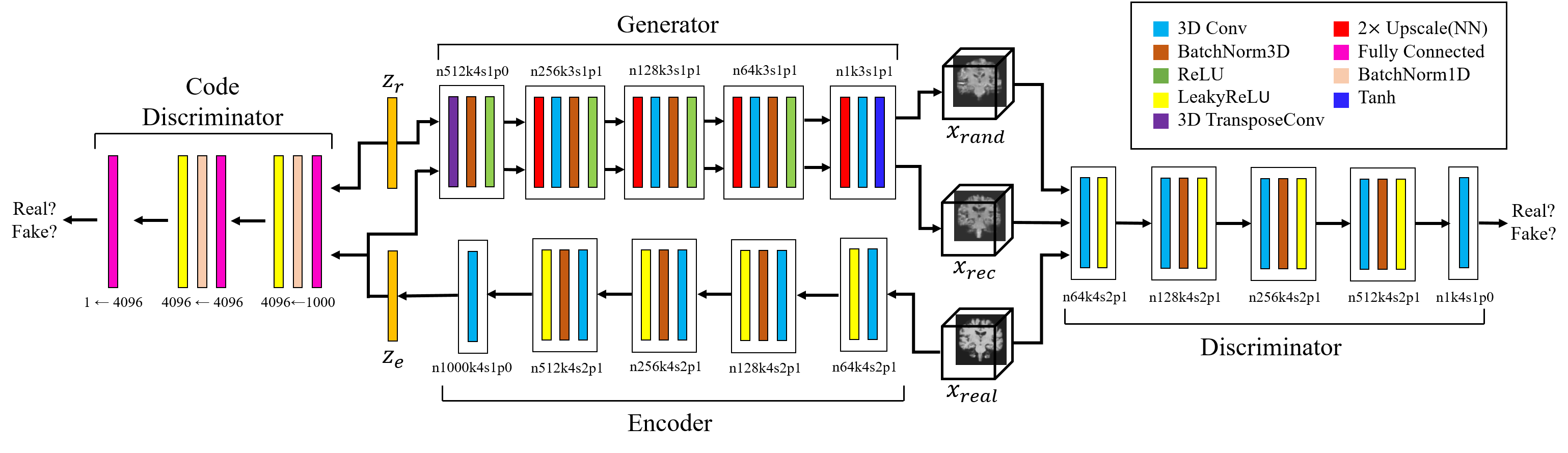}
    \caption{Detailed architecture of the proposed model. $n$, $k$, $s$, and $p$ refer to the number of the channels, the kernel size, the stride size, and the padding size, respectively. $x_{rand}$ is the generator output from random vectors $z_r$ and $x_{rec}$ is the output from encoded vectors $z_e$. }
    \label{fig:fig1}
\end{figure}

We adapt the network structure of the $\alpha$-GAN to 3-dimensional generation tasks, as shown in Figure.~\ref{fig:fig1}. Our discriminator and encoder networks have five 3D convolution layers, each of which uses 4$\times$4$\times$4 filters. Since the output from the last layer has to be a single value for the discriminator and a vector for the encoder, output channel sizes are set accordingly. We use Batch Normalization (BatchNorm) and Leaky Rectified Linear Unit (LeakyReLU) layers after each layer. In the first and the last layer, BatchNorm is removed for maintaining the originality of the input and the output. 
 
In our generator network, we use resize-convolution~\cite{Resizeconv} to reduce the number of parameters and checkerboard artifacts. Instead of transpose convolution layers, conventional nearest neighbor upscale is applied before convolution layers with 3$\times$3$\times$3 filters both. For training stability, BatchNorm and ReLU are applied after each convolution layer except for the last layer, where BatchNorm is removed and Hyperbolic tangent (Tanh) activation is used. Our code discriminator network consists of three fully-connected layers. Similar to the discriminator, LeakyReLU and BatchNorm layers are placed between each fully-connected layer. 

Since the model has difficulty learning the proper distribution with a small latent vector size of $\sim$100, we use moderately large latent vectors of size 1000 after empirically verifying that the large latent vectors have enough capacity to reflect the variety of images (See \textbf{Experiments and Results}).

\noindent
\textbf{Loss Function.} Our model with the basic GAN loss function suffers from significant instability in training, with the generated data distribution deviating from the real. Therefore, we use the loss function of WGAN~\cite{WGAN}, which measures the Wasserstein distance between both distributions. The distance metric is derived as $\mathbb{E}_{x_{gen}}[D(x_{gen})]-\mathbb{E}_{x_{real}}[D(x_{real})]$.
We also include the gradient penalty term~\cite{WGAN-GP}, which improves upon the vanilla WGAN loss by replacing the gradient clipping with following for maintaining the 1-lipschitz condition: When $\hat{x}$ is any point sampled between real and generated samples, the gradient penalty term $L_{GP}$ is described as $\mathop{\mathbb{E}}_{\hat{x}}[(||\nabla_{\hat{x}} D(\hat{x})||_2 -1)^2]$. $D$ and $x$ can be replaced into $C$ and $z$ in code discriminator.

As with $\alpha$-GAN, our loss function has a total of four loss terms, one for each network. In the discriminator loss $L_D$, the reconstructed images $G(z_e)$ and random generated images $G(z_r)$ are both treated as fake. Therefore, the discriminator loss term is the sum of two distance metrics. The loss term for code discriminator $L_C$ and encoder $L_E$ has the identical form as discriminator and generator loss of WGAN-GP, except that the input values are latent vectors $z_r$ and $z_e$ instead of images. In the case of generator loss $L_G$, reconstruction loss term is added as $L1$ distance between reconstructed images $G(z_e)$ and real images $x_{real}$. Gradient penalty terms $L_{GP-D}$ and $L_{GP-C}$ are also added to $L_D$ and $L_C$. The encoder loss $L_E$ is not specified here as it is the generator version of the code discriminator loss. For the parameters $\lambda_1$ and $\lambda_2$, we use the fixed value of 10 in both cases. Our final loss function is as follows:
\begin{equation}
L_D = \mathbb{E}_{z_e}[D(G(z_e))]+\mathbb{E}_{z_r}[D(G(z_r))]-2  \mathbb{E}_{x_{real}}[D(x_{real})]+\lambda_1L_{GP-D}  
\end{equation}
\begin{equation}
L_G =-\mathbb{E}_{z_e}[D(G(z_e))]-\mathbb{E}_{z_r}[D(G(z_r))] +\lambda_2||x_{real}-G(z_e)||_{L1}
\end{equation}
\begin{equation}
L_C = \mathbb{E}_{z_e}[C(z_e)]-\mathbb{E}_{z_r}[C(z_r)]+\lambda_1L_{GP-C}
\end{equation}

\noindent
\textbf{Training.}
In training, the encoder and the generator are considered as one network. Thus, we sum up the loss functions of the two networks as one and optimize the networks in the order of encoder-generator, discriminator, and code discriminator.  Since the generator has slower optimization speed, we update it twice in one step.  

\section{Dataset and Preprocessing}
\noindent
\textbf{Normal MRI Data.}
For the generation task, the Alzheimer's Disease Neuroimaging Initiative (ADNI)\footnote{\href{http://adni.loni.usc.edu/}{adni.loni.usc.edu}.} dataset is used. 991 T1 structural images labeled control normal (CN) are selected as real data. To prevent an extremely high variability between subjects, the non-brain areas of raw MR images are removed with recon-all function from FreeSurfer\footnote{\href{http://surfer.nmr.mgh.harvard.edu/fswiki}{surfer.nmr.mgh.harvard.edu/fswiki}.} software package. The process is done by the dataset providers. Redundant planes with all-zero values are trimmed, then images are resized into 64$\times$64$\times$64. To compensate for the lack of samples without transforming the attributes of the original data, only left-right flip and intensity randomization are applied.

\noindent
\textbf{Diseased MRI Data.}
For additional experiments, two brain disorder MRI datasets are used. First, we train the model with BRATS 2018~\cite{BRATS2,BRATS1} dataset for brain tumor MRI generation, using 210 subjects in the training dataset labeled as 'HGG\footnote{high-grade gliomas}.' Each subject has images of four different modalities (T1, T1ce, T2, FLAIR). We use FLAIR and T2, which have imminent visual features of tumors. Second, Anatomical Tracings of Lesions After Stroke (ATLAS)\footnote{\href{http://fcon_1000.projects.nitrc.org/indi/retro/atlas.html}{fcon\_1000.projects.nitrc.org/indi/retro/atlas.html}.}~\cite{ATLAS} dataset is used for stroke MRI generation. The dataset contains 220 T1w images, which have diverse stroke lesions. Non-brain areas are removed using segmentation tools from SPM12\footnote{\href{https://www.fil.ion.ucl.ac.uk/spm/software/spm12/}{www.fil.ion.ucl.ac.uk/spm/software/spm12}.}. 
Images from both datasets are processed with redundant planes trimming, resizing into 64$\times$64$\times$64, and left-right flip.

\section{Experiments and Results}
\noindent
\textbf{Experiment Details.}
Our experiments are conducted on an NVIDIA Titan Xp 12GB GPU. Programs are implemented with Python, using the Pytorch deep learning library\footnote{Our code for the experiment and the models is available at \url{https://github.com/cyclomon/3dbraingen}}. For training the model, the Adam optimizer is used with learning rate of 0.0002 for all four networks, and the size of mini-batch is set to 4. We normalize all data into the range [-1, 1]. All the generated samples are generator outputs from random latent vectors $z_r$. The results are obtained by considering 3D training images with additional augmentations (flip, intensity randomization) as real.

\noindent
\textbf{Baseline Models.}
For comparison, 3D-$\alpha$-GAN,  3D-VAE-GAN, 3D-WGAN-GP are used as baseline models. The VAE-GAN~\cite{VAEGAN} model has similar motivation and structure to our model in that it combines VAE and GAN for training stabilization and image blurriness. WGAN-GP is chosen to evaluate effectiveness of the proposed approach in alleviating the mode collapse problem. Lastly, we include the baseline 3D-$\alpha$-GAN model trained with the basic GAN loss to verify the validity of our WGAN-GP loss function. All models are composed of 3D-convolutional layers and implemented with 1000-dimensional latent vectors as input.

\begin{figure}[t]
    \centering
    \includegraphics[width=0.9\linewidth]{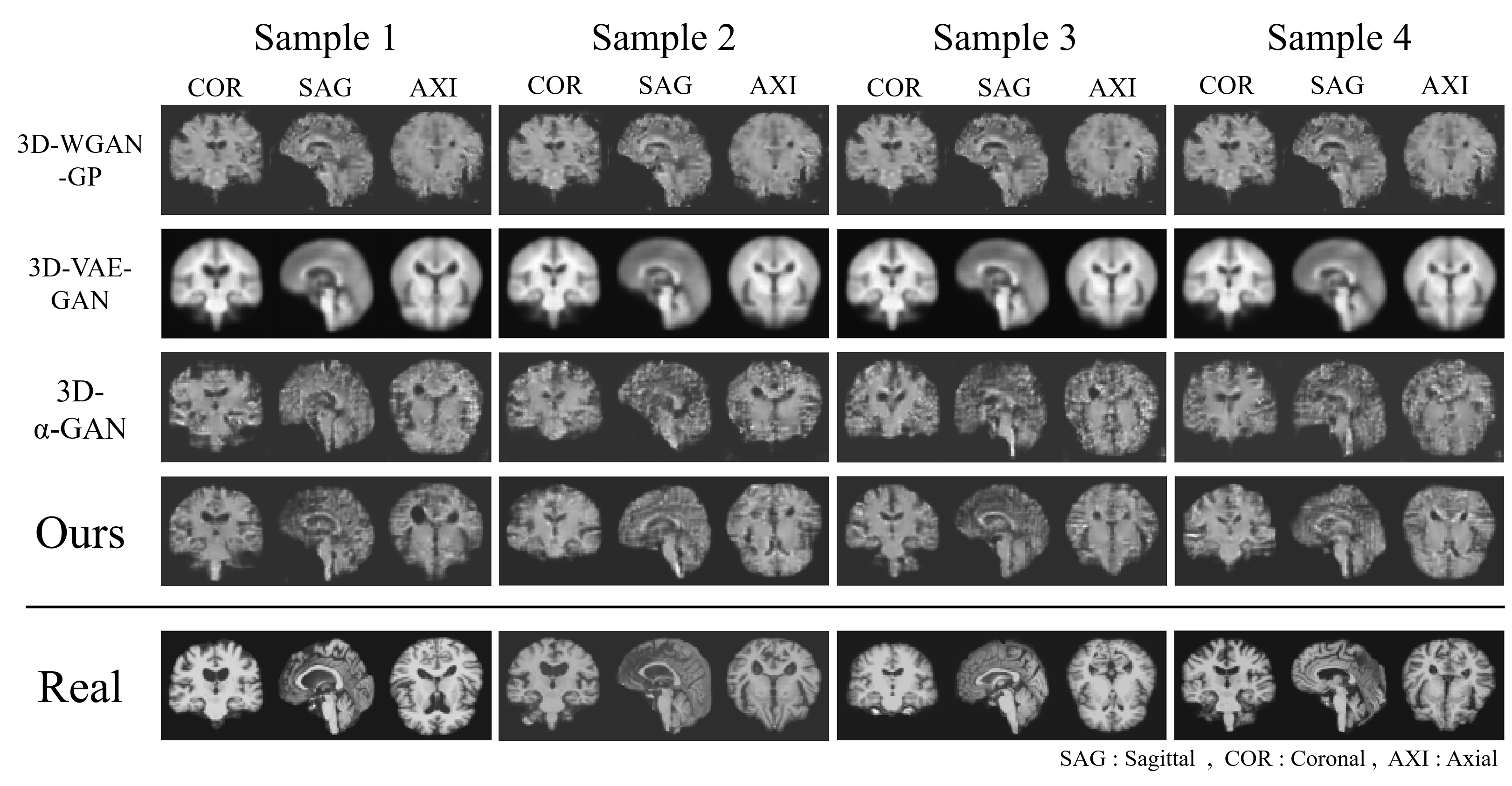}
    \caption{Real and generated samples of normal brain MRI.}
    \label{fig:fig2}
\end{figure}

\noindent
\textbf{Generated Images.}
In Figure.~\ref{fig:fig2}, center-cut slices of generated 3D samples are shown. The generated images from 3D-WGAN-GP and 3D-VAE-GAN are extremely similar to each other, which indicates that the models have mode collapse. Samples from 3D-VAE-GAN are blurry and the detailed features of the brain disappear. In the 3D-WGAN-GP and baseline 3D-$\alpha$-GAN, images suffer from unwanted artifacts. In contrast, samples from our model reflect the detailed attributes of brains (e.g. sulci, gyri) with proper diversity. More samples are provided as full volume slices in \textbf{Supplementary Material}.

\begin{table}
\centering
    \caption{Quantitative results}\label{tab:tab1}
    \begin{tabular}{l V{2.5} c|c}
     &MMD$\times10^{-4}$&MS-SSIM\\\hlineB{2.5}
    3D-WGAN-GP&0.327&0.996\\
    3D-VAE-GAN&0.075&0.972\\
    3D-$\alpha$-GAN&0.131&\textbf{0.843}\\\hline
    Ours-$z$100&0.211&0.927\\
    Ours-$z$1000 (Best)&\textbf{0.072} &0.829\\
    Ours-$z$2048&0.108&0.867\\\hline
    Real & - & 0.846\\
    \end{tabular}
\end{table}
    
\noindent
\textbf{Quantitative Results.}
Table~\ref{tab:tab1} shows the quantitative results. First, we investigate the distribution distance between the real and generated samples with maximum mean discrepancy (MMD) scores~\cite{MMD}. Due to the memory issue, we use the averaged batch-wise MMD$^2$ calculations for the entire data. The final MMD scores are the averaged values of 100 tests. We use batch size of $B = 8$. If a flattened batch of generated samples is set as $g$, and real samples as $r$, a single batch-wise MMD$^2$ value is calculated as: $\frac{1}{B^2}\sum g\cdot g^\intercal  + r \cdot r^\intercal - 2g \cdot r^\intercal$.

The generated samples from our model has the lowest MMD scores, indicating that the distribution of our result samples are the closest to that of the real data.
Second, we evaluate the generation diversity with multi-scale structural similarity metric (MS-SSIM)~\cite{Odena_2017}. MS-SSIM scores are calculated with the average from 1000 sample pairs. In the case of 3D-WGAN-GP and 3D-VAE-GAN, the similarity scores between generated samples are very large, thus mode collapse occurs. Baseline $\alpha$-GAN and our models can generate diverse samples with relatively similar scores to that of the real data. However, the model with too small latent vectors fails to escape the mode collapse. 

\begin{figure}[!t]
    \centering
    \adjustbox{max width=0.9\textwidth}{
    \subfloat[\label{fig:3a}]{\includegraphics[width=0.33\textwidth]{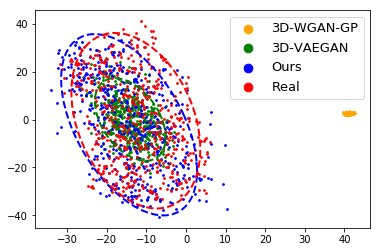}}\hfill
    \subfloat[\label{fig:3b}]{\includegraphics[width=0.33\textwidth]{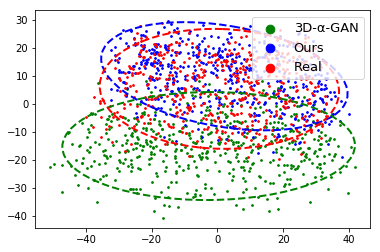}}\hfill
    \subfloat[\label{fig:3c}]{\includegraphics[width=0.33\textwidth]{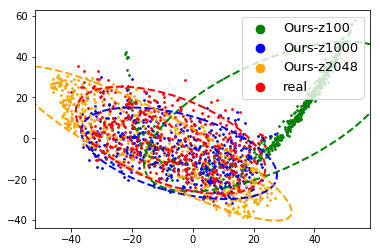}}}
    \caption{PCA results of generated samples. (a) Samples from real data, 3D-WGAN-GP, 3D-VAE-GAN, Ours (3D-$\alpha$-WGAN-GP). (b) Samples from real data, 3D-$\alpha$-GAN, Ours (3D-$\alpha$-WGAN-GP). (c) Samples with changing latent vector sizes. }\label{fig:fig3}
\end{figure}

\noindent
\textbf{Qualitative Results.}
To qualitatively analyze the results, we compare the generated and real data distributions by visualizing 512 generated samples from each model and 512 randomly selected samples from real data with Principal Component Analysis (PCA).
Figure.~\ref{fig:fig3}(a,b) shows the scatter plot results of PCA, with each point representing a data sample. In the case of 3D-WGAN-GP, only a small range of images are generated and the distance from other distributions is very large. With the 3D-VAE-GAN, the output samples are within the real data distribution range, but it produces a limited variety of images. With the baseline 3D-$\alpha$-GAN model, the mode collapse problem is solved, but the distribution is still not closely matched to the real data. Compared to the baseline models, the results from our proposed model have the most similar distribution range to the real data. 
The results show that $\alpha$-GAN, but not VAE-GAN, exceeds the threshold of mode collapse because $\alpha$-GAN accomplishes a more accurate latent distribution matching by replacing KL divergence with JS divergence, which can measure the exact distance between the distributions.

Figure.~\ref{fig:fig3}(c) shows the effects of different latent vector sizes. Experiments are carried out with our model, varying the latent variable size. The figure shows that the model with small latent vector size of 100 suffers from severe mode collapse. While it is possible to obtain acceptable results using a very large size, many samples are out of the real data distribution. The best results can be obtained using a moderately large latent vector size, corroborating the quantitative results.

\begin{figure}[t]
    \centering
    \includegraphics[width=0.9\linewidth]{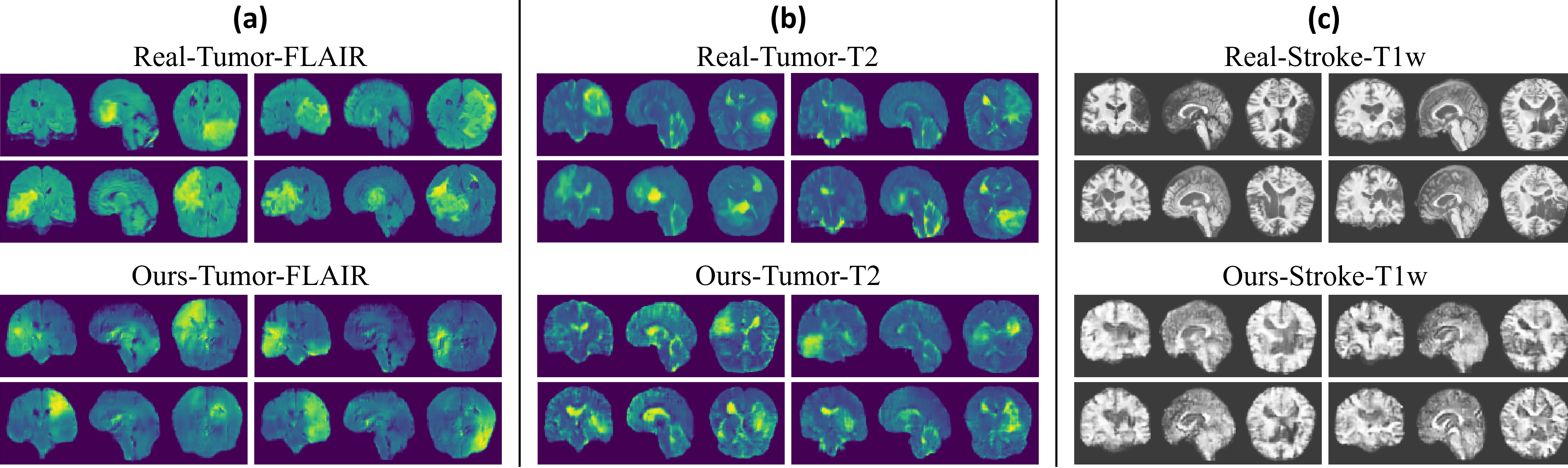}
    \caption{Real and generated samples of diseased brain MRI.}
    \label{fig:fig4}
\end{figure}
\noindent
\textbf{Diseased MRI generation.} To test the versatility of our model, we train the model with brain tumor and stroke 3D MRI data. We use three different training sets: Tumor-T2, Tumor-FLAIR and Stroke-T1w. Figure.~\ref{fig:fig4}(a,b) shows image samples of severe cases of brain tumor in FLAIR and T2. For better visualization, we display color-mapped images where yellow indicates higher and blue indicates lower intensity. Our model generates realistic samples with brain tumor lesions at various positions while properly reflecting the characteristics of different modalities. 
Figure.~\ref{fig:fig4}(c) demonstrates the generation results of T1w brain MRI with large stroke lesions. Although the features of stroke are totally different from tumor, our model can generate plausible samples with damaged areas in various positions.
Generated images from other models are severely deteriorated (See \textbf{Supplementary Material}).  

\section{Conclusion}
In this paper, we present a novel auto-encoder based GAN model that generates realistic 3D brain MRI data with a small amount of training samples. We demonstrate that our model outperforms alternative structures in capturing the real data distribution and generating diverse samples. Moreover, our model can be applied to various types of data and generates images that accurately reflect the attributes of each type. The results suggest that our model can be used for efficient data augmentation of 3D brain MRI data, opening up possibilities for applications such as disease diagnosis.

As future work, we are generating disease images that are difficult to generate with image-to-image translation (e.g. Alzheimer's) and planning to show the improvement in diagnostic performance with the generated images.

\subsubsection{Acknowledgements:}
This work was supported by Institute for Information \& communications Technology Promotion(IITP) grant funded by the Korea government(MSIT) (No.2016-0-00563, Research on Adaptive Machine Learning Technology Development for Intelligent Autonomous Digital Companion), and Engineering Research Center of Excellence (ERC) Program supported by National Research Foundation (NRF), Korean Ministry of Science \& ICT (MSIT) (GrantNo. NRF-2017R1A5A1014708).

\bibliographystyle{splncs04}

\includepdf[pages=-]{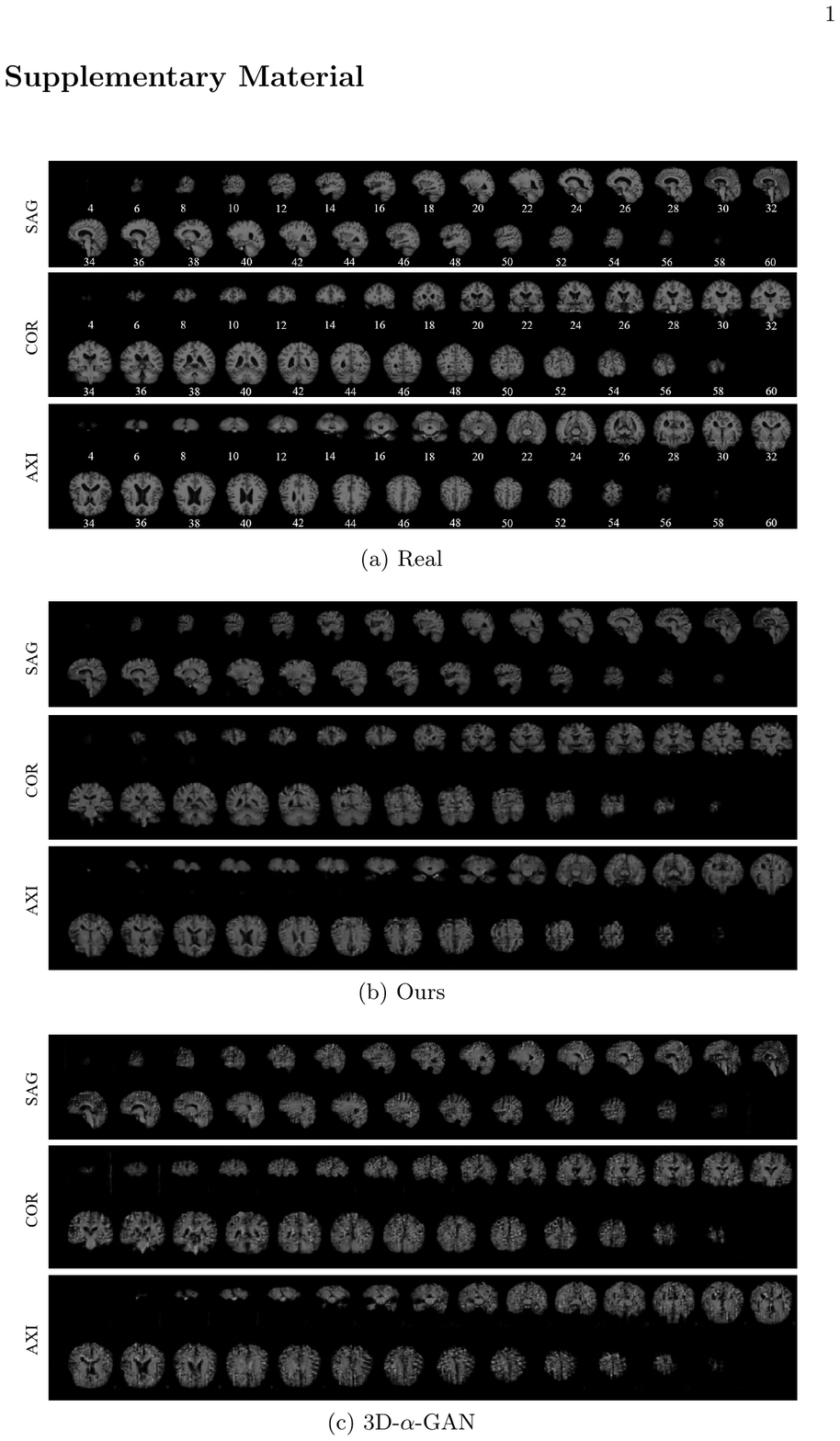}
\end{document}